\documentclass[12pt]{article}
\usepackage{graphicx}
\usepackage{bm}
\usepackage{amssymb}
\begin{document}
\begin{titlepage}


\vspace{1cm}

\begin{flushright}
{\bf Budker INP 2002-14\\
February 27, 2002 }
\end{flushright}
\vspace{2cm}



\begin{center}{\Large {\bf Comparison of theory with experiment \\for positron production
from high-energy \\electrons moving along crystal axes }}\\ 
\vspace{1.0cm}
{\Large V.N.Baier and V.M.Strakhovenko} \\
G.I. Budker Institute of Nuclear Physics,\\
630090 Novosibirsk, Russia\\
\end{center}
\vspace{4.0cm}


\begin{abstract}

Various positron distributions are obtained using an approach developed
earlier for the description of the electron~-~photon showers in axially aligned single crystals. Based on these distributions, characteristics of the positron yield measured in recent experiments are calculated. Theoretical estimations display a rather good agreement with experimental results obtained using 3 to 10~GeV electrons aligned to
the $<111>$~-~axis of the tungsten crystals.

\end{abstract}
\end{titlepage}
\newpage
\section{Introduction}
\label{}

An efficient positron source is one of the important components of
future electron~-~positron colliders. Positrons are generated from
electrons in the course of the $e^{-}e^{+}\gamma$~-~shower developing in a
medium. In high-energy region, the basic processes involved in the shower
development are typically considerably enhanced in oriented crystals as compared with
corresponding amorphous media . The most pronounced effects
take place at axial alignment when initial electrons are moving along the main axes of a crystal. This alignment alone will be considered below. Then, according to \cite{BKS1},the radiation intensity in a crystal exceeds that of the conventional bremsstrahlung starting
with electron energies $\varepsilon \sim $ 1 GeV. Simple estimations of the width of the power spectrum performed in \cite{BKS1} indicate a  soft character of this spectrum.
Basing on these properties of the photon emission process,
 the use of this phenomenon in positron source for
future accelerators was proposed \cite{BCh}. The pair production rate
which is due to the coherent (crystal) effects exceeds that of the
standard (Bethe-Heitler) mechanism starting with photon energies $\omega
\simeq \omega_{th}$. The value
of $\omega_{th}$ is about 22~GeV for the $ <111> $~-~axis of tungsten being several
times larger for another crystals. (See review \cite{BKS2} and recent book \cite{BKS2a} for
further details concerning QED~-~processes in crystals.) For energies well above $\omega_{th}$, the crystal effects become
really  strong and may be used to create
effective and compact electromagnetic calorimeters \cite{BKS2b}.
For very high energies ($\varepsilon \gg \omega_{th}$) of initial and created particles,
 kinetic equations describing the
shower development were solved analytically \cite{BKS3}. Though
the initial electron energies were high enough in the first
experimental investigation \cite{Rmed} of shower formation in
crystals, energies of detected particles were too low to allow
us the direct comparison with \cite{BKS3}. To explain the results
of \cite{Rmed}, Monte-Carlo simulations were performed in
\cite{BKS55}. The probabilities of basic processes used in
\cite{BKS55} were obtained within so-called constant field approximation. A
good agreement was demonstrated in \cite{BKS55} with the results of \cite{Rmed}
for Ge crystals.

When the initial electron energy is below $\omega_{th}$,
photons are mainly  emitted with energies $\omega \ll \omega_{th}$
and so, up to minor modifications (see \cite{BKS88}, \cite{BS1}), the
pair production process proceeds in a crystal as in an amorphous
medium. The enhancement of radiation from initial electrons is thereby the main crystal effect in this energy region. A substantial advance in the description of shower
formation at axial alignment was caused by the invention
of the semi~-~phenomenological radiation spectrum \cite{BKS4} . This
allows one to consider the relatively low (of a few GeV) energy
range of the initial electrons which is presumed for the efficient
positron source. The radiation
intensity  increases with the initial electron
energy. As a result, at some energy the effective radiation
length $L_{ef}$ in a crystal becomes smaller than the conventional
radiation length $L_{rad}$ and continues its decrease at further
increase of the energy. All numerical examples will be given below for
the electron beam aligned with the $<111>$~-~axis of the tungsten crystals.
Then we have for the quantity $L_{ef}$ defined as in Sec.3 of
\cite{BKS4}: $L_{ef}$(1~Gev) $\simeq$ 0.166 cm, $L_{ef}$(4~Gev)
$\simeq$ 0.084 cm, and $L_{ef}$(8~Gev) $\simeq$ 0.061 cm. In
a hybrid target which consists of the crystal part followed by the amorphous one,
a thickness of the crystal constituent of several $L_{ef}$ is obviously quite enough.
Indeed, at the depth  $L_0 \approx (3 \div 4) L_{ef}$ most of the particles,
including the initial electrons, are sufficiently soft to reduce the
coherent contribution to the radiation to the level of the
incoherent one. Thereby, the further development of the shower
proceeds more or less in the same way for the crystal or amorphous
type of the remaining part of a target. We emphasize that the
crystal part $L \leq L_0$ of a target serves as a radiator, and
secondary charged particles are still not so numerous at this
stage of the shower development. Therefore only a small portion
of the total energy loss is deposited in the crystal part of a
target which considerably reduces a danger of its overheating. The softness of
photon spectra is another important feature of the crystal
radiator giving additional advantages for the positron production
in comparison with the entirely amorphous target. To get more definite
idea concerning a shape of the power spectrum one can use its explicit
form given by Eq.(2) in \cite{BKS4}. To present the scale, let us
list some values $\omega_{max}$ where this spectrum is maximum: $\omega_{max}(1 GeV) \simeq $31~MeV, $\omega_{max}(4 GeV) \simeq $170~MeV, and $\omega_{max}(8 GeV) \simeq $490~MeV.
 Note that a width of the spectrum is typically several times larger than
$\omega_{max}$. The increase in the number of relatively soft
photons turns out to be much more pronounced than that in the
total radiation intensity. In the end, just this fact leads to
the substantial enhancement of the positron yield from crystal
targets.

Recently the positron production in axially aligned  single crystals was
studied in two series of experiments performed at CERN \cite{CBA00}, \cite{C01}
and KEK \cite{ITY00}, \cite{O01}. The initial energy of electrons was 3~GeV
\cite{ITY00}, 6 and 10~GeV \cite{C01}, 8~GeV \cite{O01}, and 10~GeV
\cite{CBA00}. In all cases the initial electron beam was aligned with the
$<111>$~-~axis of a tungsten crystal that sometimes served as the crystal part
of a hybrid target which contained an additional amorphous tungsten target. A
noticeable enhancement of the low-energy  positron yield was observed in all
experiments cited above when the yield from a crystal target  was compared with
that from an amorphous target of the same thickness. The experimental results
and our theoretical estimations presented in the next Section display a rather
good agreement with each other.

\section{Comparison of theory with experiment}

Theoretical results for the conditions of the experiments cited above were
obtained using the approach developed in  \cite{BKS4} and \cite{BS1} where
various positron and photon distributions as well as deposited energies in
different crystals were calculated  for the energy range of initial electrons
from 2 to 300~GeV. In these papers, all the formulas used in  Monte-Carlo
simulations of the specific $e^{-}e^{+}\gamma$~-~shower characteristics are given
in the explicit form. Remember that our simplified description of the shower
development takes into account coherent ( induced by the regular motion of
particles in the field of crystal axes ) and incoherent ( like that in an
amorphous medium ) mechanisms of photon emission and pair production processes.
The multiple scattering and the ionization energy loss of electrons and
positrons are taken into account neglecting crystal effects. The coherent
radiation from channelling and moving not very high above the axis potential
barrier particles is described using the semi~-~phenomenological spectrum
suggested in \cite{BKS4}. A corresponding computer code was developed. This
allows one to calculate energy, angular, and coordinate distributions of
positrons emergent from a crystal or hybrid target and to find an amount of the
energy deposition. We think that the investigation of such
distributions should be the main object of the experiments having a creation of
the crystal assisted positron source as their ultimate aim.

\subsection{Experiment (CERN) at $ \varepsilon_0 = 10$~GeV }

\begin{figure}[h]
  \centering
  \includegraphics[width=0.48\textwidth
  ]{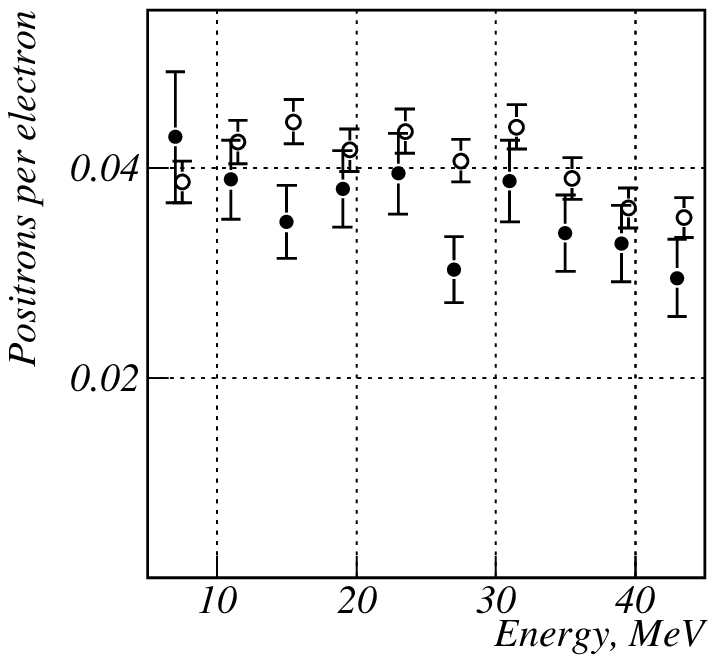}\hspace{0.025\textwidth}
 \includegraphics[width=0.48\textwidth
 ]{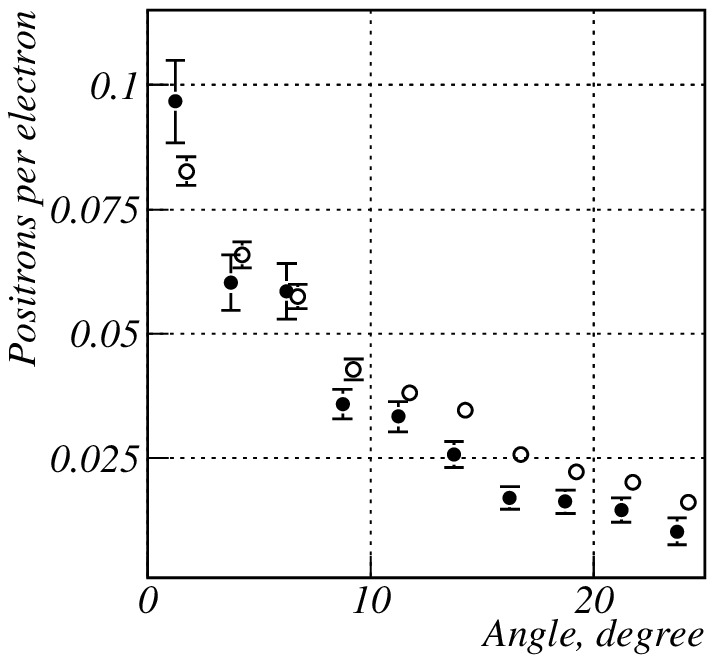}
 \caption{Spectral (left) and angular (right) distributions of positrons from
10~GeV electrons traversing 8~-~mm~-~thick crystal tungsten target along the
$<111>$-~axis. Open circles - simulation, filled circles - experiment.
}\label{Fig:Cern}
 \end{figure}

   Among experiments cited above, spectral~-~angular distributions of created
positrons were measured only in WA103 experiment at CERN ( see \cite{CBA00},
\cite{C01}) where our code was used in simulations as the event
generator. This simulation allowed for the acceptance conditions and the
efficiency of the detectors used. Shown in Fig.\ref {Fig:Cern} taken from
\cite{C01} is one example of the measured and simulated distributions of positrons from
10~-~GeV electrons aligned with the $<111>$-~axis of the 8~-~mm~-~thick crystal
tungsten.
The angular acceptance conditions in WA103 experiment were approximately
$|\vartheta_{V}^{out}|\leq 1.5^{\circ} $ for the vertical and $0\leq
\vartheta_{H}^{out}\leq 25^{\circ}$ for the horizontal angle of outgoing positron
with respect to the initial electron beam direction. We shell see below that a
shape of the positron spectrum depends on a degree of collimation.
 The one-dimensional ( over $\vartheta_{H}^{out}$) angular
distribution is presented for positrons having energies in the 5$\div$45~MeV
range. We emphasize that a relative difference between measured and simulated
results typically does not exceed 20~$\%$ in both spectral and angular
distributions as seen in Fig.\ref {Fig:Cern}. We are aware that preliminary
results for another settings used in the same experiment do not contradict with
the estimated scale of the difference between the data and theoretical
predictions. We hope that this interrelation will not become worse after
performing the complete analysis of the data  which now is underway. This
analysis will also give more detailed information concerning spectral~-~angular
distributions of positrons depending on initial electron energies and target thicknesses.

\subsection{Experiment (KEK) at $ \varepsilon_0 = 3$~GeV }

The main goal of the experiment \cite{ITY00} was an attempt to apply the
crystal target to a working electron/positron linac, the injector for the
electron~-~positron collider B~-~Factory at KEK. Thus, the acceptance
conditions for created positrons were determined by the momentum acceptance of
the positron linac with a matching section which is 8.2~MeV/c $ < p <$
11.6~MeV/c and $p_{\perp} <$ 2.4~MeV/c. The hybrid target used consists of
1.7~-~mm~-~thick tungsten crystal followed by 7~-~mm~-~thick amorphous
tungsten. The observed positron yield was enhanced by the factor 1.40 when the
$<111>$ crystal axis was aligned with 3~GeV incident electron beam as compared
to the case of the disoriented crystal. Our number for this enhancement is 1.47
being only 5~$\%$ larger than the experimental one. Note that in the experiment
\cite{ITY00} the crystal and amorphous parts of the hybrid target were
separated by the distance of 70~mm. This circumstance, which, in principle, may
slightly change the enhancement value, was not taken into account in our
calculation. Recollect that an amount of the energy deposited in the crystal part
($ \varepsilon_{dep}^{cr}$ ) of a hybrid target may be much smaller than that
($ \varepsilon_{dep}^{am}$) in the amorphous one. Such interrelation of $
\varepsilon_{dep}^{cr}$  and $ \varepsilon_{dep}^{am}$ should take place in
the case of \cite{ITY00}, where the crystal thickness is about 1.8~$ L_{ef}$ ( see discussion in the Introduction ).
This is confirmed by our calculations which give $ \varepsilon_{dep}^{cr} \simeq $
11~MeV and $ \varepsilon_{dep}^{am} \simeq$ 277~MeV per one incident electron.

\subsection{Qualitative features of positron distributions and experiment (KEK) at
$ \varepsilon_0 = 8$~GeV }

In \cite{O01} the positron production efficiency
from 2.2~-~mm, 5.3~-~mm and 9.0~-~mm~-~thick tungsten crystals was measured
using an 8~-~GeV electron beam. Positrons produced in the forward direction with
momenta 10, 15 and 20 ~MeV/c were detected by a magnetic spectrometer. Thus,
only several points in the energy distribution were determined under hard
collimation conditions. Therefore, before going on to the comparison of the
experimental results with our, let us remind some important qualitative
features of spectral~-~angular distributions using 8~GeV electrons and the
$<111>$~-~axis of the tungsten crystals as an example. For the sake of comparison,
the corresponding distributions for amorphous tungsten will be presented as
well. Below all the quantities characterizing a positron yield are normalized
per one incident electron.

\begin{figure}[h]
  \centering
  \includegraphics[width=0.48\textwidth
  ]{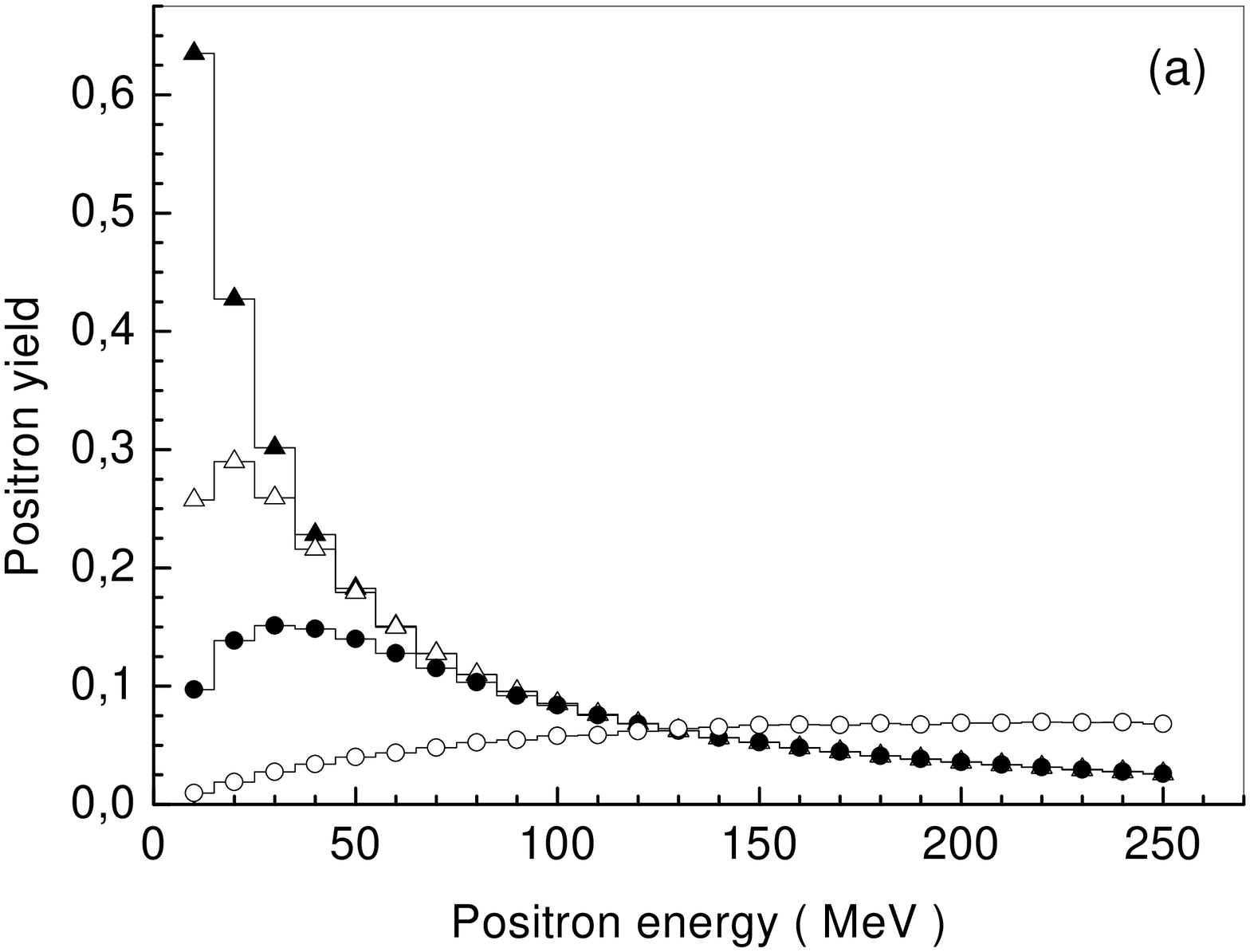}\hspace{0.025\textwidth}
 \includegraphics[width=0.48\textwidth
 ]{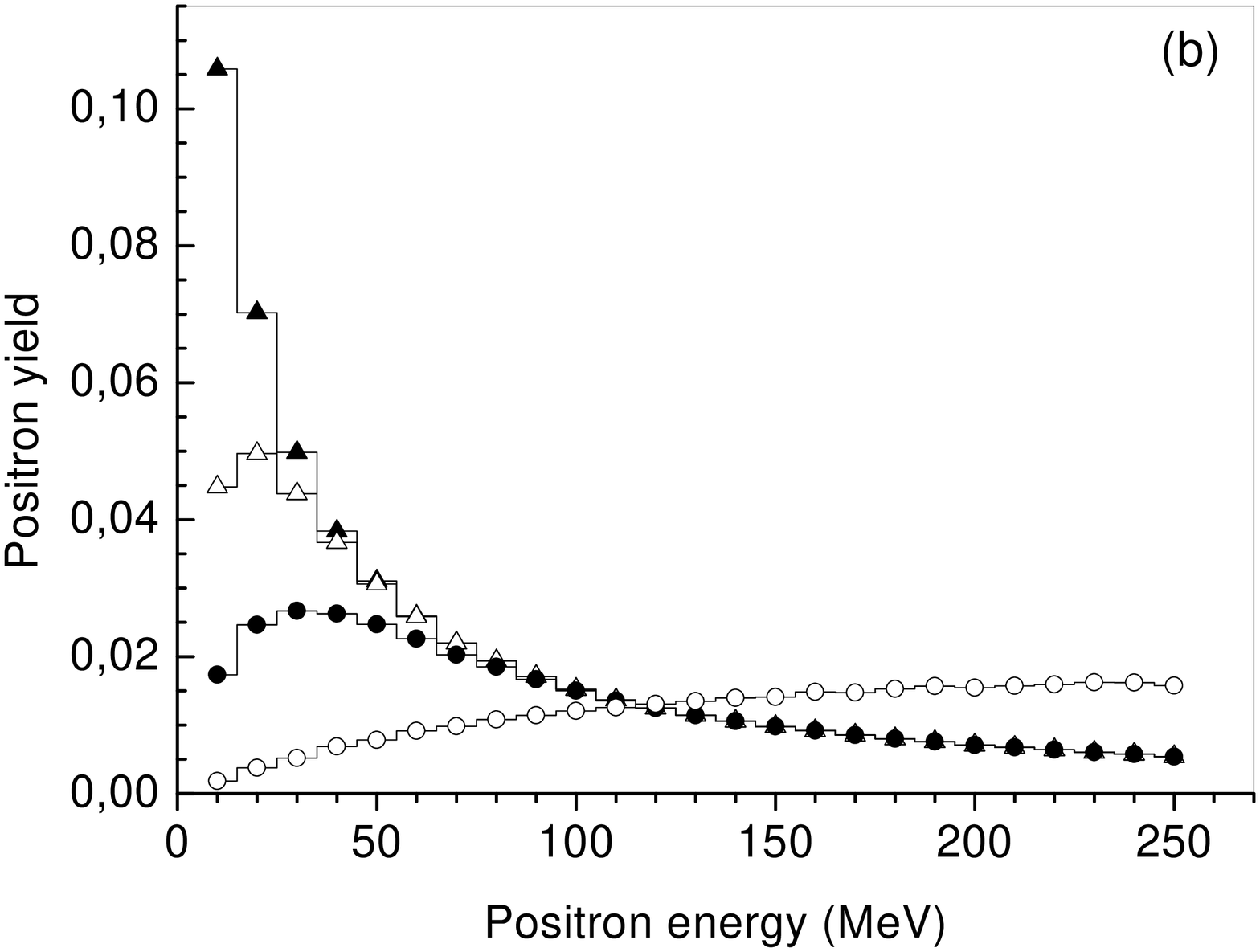}
 \caption{Positron yield  depending on energy from  2.2~-~mm~-~thick crystal
 (a) and amorphous (b) targets at different collimation
. Filled triangles - no collimation ($\vartheta_{out}\leq 180^{\circ}$), open
triangles - $\vartheta_{out}\leq 24^{\circ}$, filled circles -
$\vartheta_{out}\leq 12^{\circ}$, and open circles - $\vartheta_{out}\leq
1^{\circ}$ (multiplied by 10). } \label{Fig:Spec2}
 \end{figure}

The use of matching systems implies some collimation ( typically
$\vartheta_{out}\leq 25^{\circ}$ ) of outgoing positrons. Shown in Fig.\ref
{Fig:Spec2} is the energy dependence ( energy step is equal to 10~MeV ) of the
positron yield  from crystal (a) and amorphous (b) targets of the same
thickness $L=2.2$~mm. In the case of the hard collimation, when
$\vartheta_{out}\leq 1^{\circ}$ ( open circles ), the yield is multiplied by 10
to make it visible. The larger a positron energy, the smaller is a typical
value of $\vartheta_{out}$ since both production and multiple scattering
processes are characterized by smaller angles for higher energies. This is seen
in Fig.\ref{Fig:Spec2}~(a) where non-collimated spectrum joins that for
$\vartheta_{out}\leq 24^{\circ}$ at $\varepsilon_{cr}^{(1)} \simeq $ 55~MeV.
The latter, in turn, joins the spectrum for $\vartheta_{out}\leq
12^{\circ}$ at $\varepsilon_{cr}^{(2)} \simeq $ 110~MeV. Such behavior is also
seen in Fig.\ref{Fig:Spec2}~(b) for the amorphous target where
$\varepsilon_{am}^{(1)} \simeq $ 50~MeV and $\varepsilon_{am}^{(2)} \simeq $
105~MeV.
In other words, positrons with energies $\varepsilon > \varepsilon^{(1)}$  are
practically concentrated within a cone $\vartheta_{out}\leq 24^{\circ}$ and
those with $\varepsilon > \varepsilon^{(2)}$ have $\vartheta_{out}\leq
12^{\circ}$. In accordance with this picture, the spectral maximum is shifted
to the right while a width of the distribution increases when the collimation
angle decreases. The enhancement $\mu$, being bin-by-bin ratio of the positron
yield from a crystal target to that from an amorphous one at the same
collimation, is almost constant for $\varepsilon < $ 45~MeV and monotonically
decreases with growing positron energy. This means that positron spectra from a
crystal target are softer. Somewhat lower values of   $\varepsilon^{(1)},
\varepsilon^{(2)}$ in the amorphous case point at the same feature. For given
collimation, a variation of the enhancement is about 20~$\%$ over the whole
energy interval presented in Fig.\ref{Fig:Spec2}. The maximum values of the
enhancement at different collimation are $\mu_{max}(\vartheta_{out}\leq
180^{\circ})\simeq 6.09 $, $\mu_{max}(\vartheta_{out}\leq 24^{\circ})\simeq
5.92 $, $\mu_{max}(\vartheta_{out}\leq 12^{\circ})\simeq 5.67 $, and
$\mu_{max}(\vartheta_{out}\leq 1^{\circ})\simeq 5.29 $. Apparently, they
diminish as a collimation angle does so.
  \begin{figure}[h]
  \centering
  \includegraphics[width=0.48\textwidth
  ]{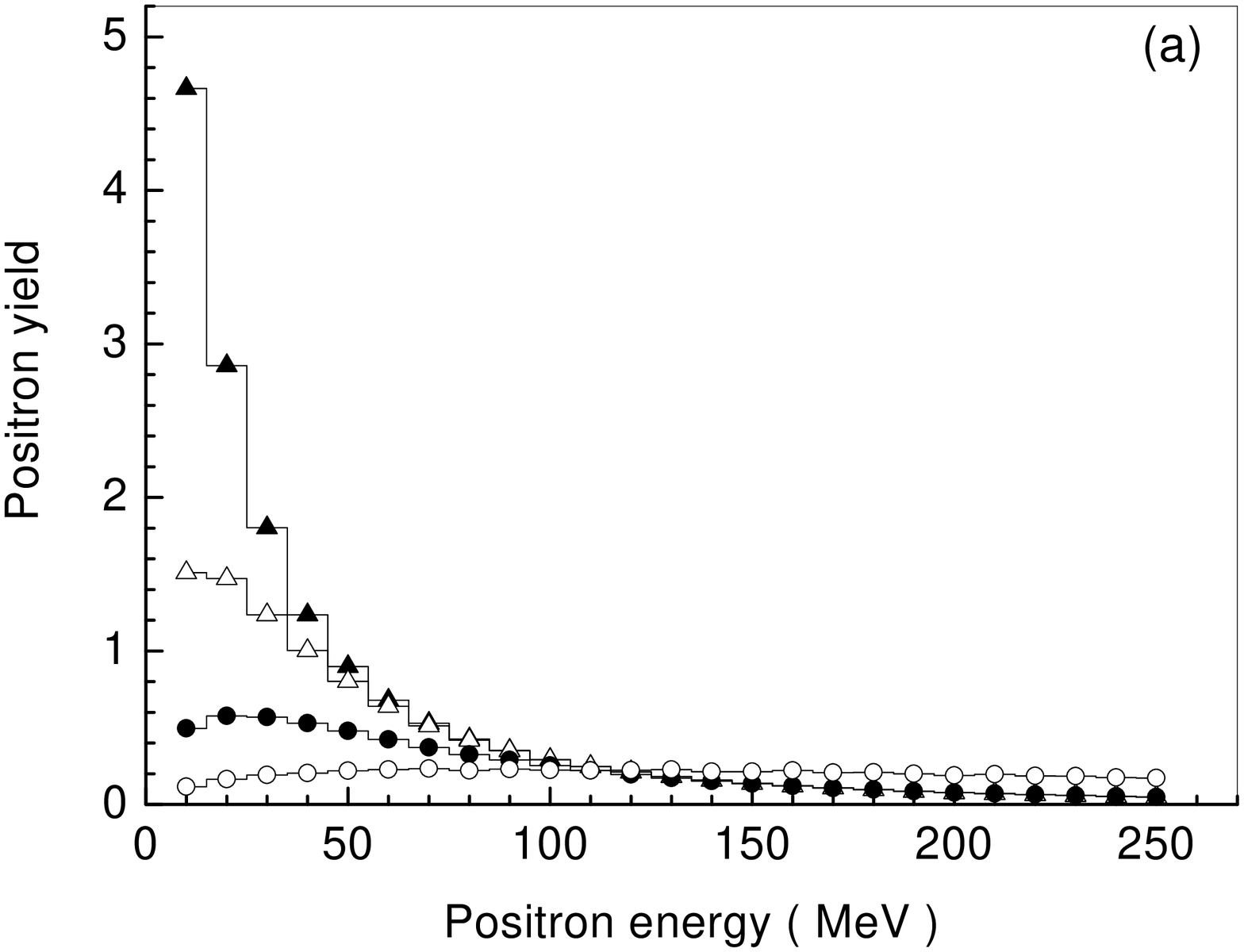}\hspace{0.025\textwidth}
 \includegraphics[width=0.48\textwidth
 ]{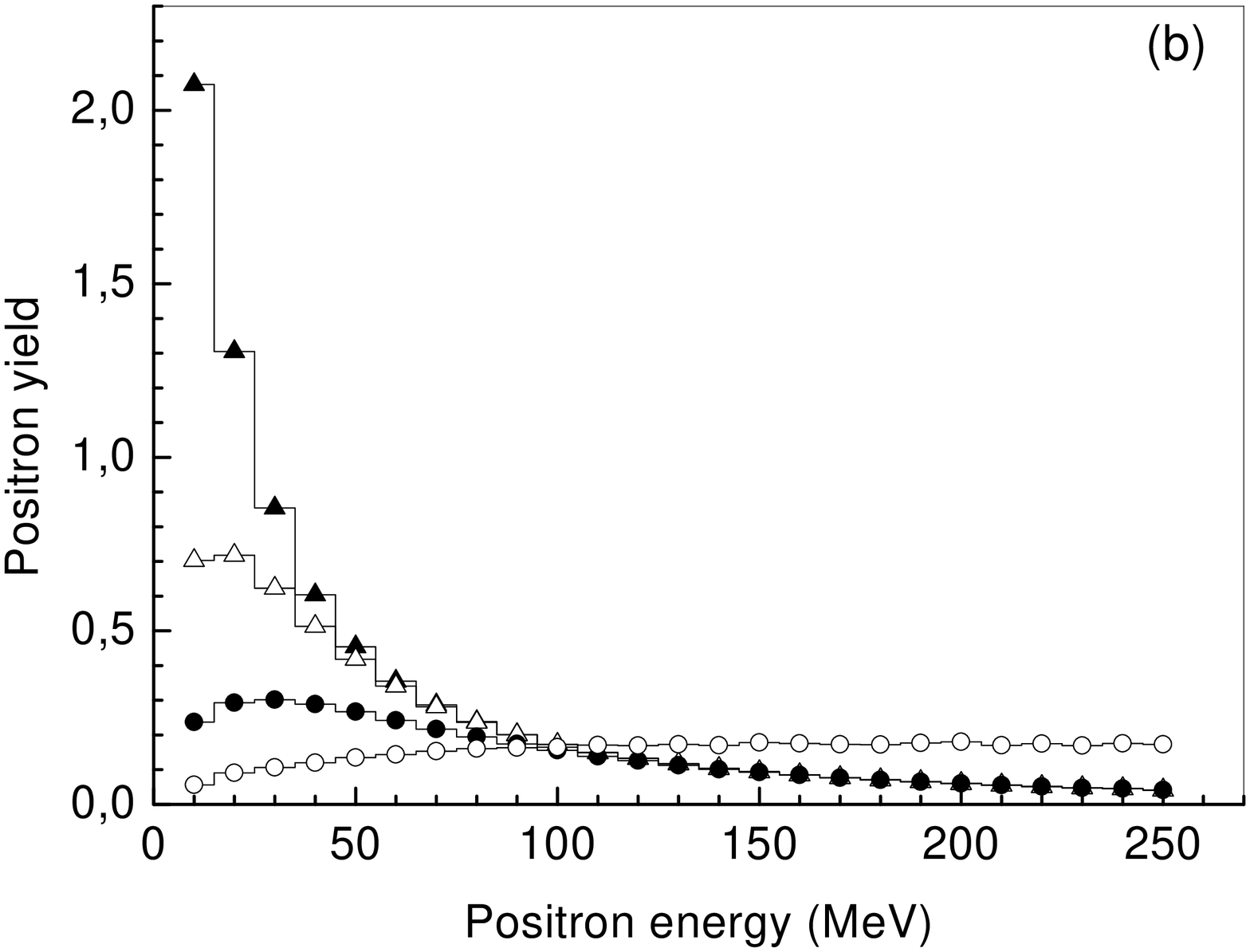}
 \caption{Positron yield  depending on energy from  9.0~-~mm~-~thick crystal
 (a) and amorphous (b) targets at different collimation
. Filled triangles - no collimation ($\vartheta_{out}\leq 180^{\circ}$), open
triangles - $\vartheta_{out}\leq 24^{\circ}$, filled circles -
$\vartheta_{out}\leq 12^{\circ}$, and open circles - $\vartheta_{out}\leq
1^{\circ}$ (multiplied by 30). } \label{Fig:Spec9}
 \end{figure}
Shown in Fig.\ref{Fig:Spec9} is the same as in Fig.\ref{Fig:Spec2} but for the
target thickness $L=9.0$~mm. The yield at $\vartheta_{out}\leq 1^{\circ}$ (
open circles ) is multiplied now by 30. A qualitative behavior of spectra
depending on the collimation angle at $L=9.0$~mm is the same as at $L=2.2$~mm.
However, all the spectra become softer for the larger target thickness. This is
indicated already by the increase in $\varepsilon^{(1)}$, $\varepsilon^{(2)}$ values which are now
$\varepsilon_{cr}^{(1)} \simeq $ 85~MeV, $\varepsilon_{cr}^{(2)} \simeq $
185~MeV, $\varepsilon_{am}^{(1)} \simeq $ 75~MeV, $\varepsilon_{am}^{(2)}
\simeq $ 165~MeV. It is clear that the magnitude of the yield from the thicker
target is essentially larger but this increase is different in the crystal and
amorphous cases. For example, in the energy range $\varepsilon < $ 45~MeV the
yield is increased by 6~$\div$~7 times for a crystal and by 17~$\div$~20 times
for amorphous samples. As a result, the enhancement at $L=9.0$~mm is almost 3
times less than at $L=2.2$~mm in this energy range.  At $L=9.0$~mm the
enhancement is peaked in the first bin ( $\varepsilon \in (5 \div 15)$~MeV) for
every collimation. Its maximum values are $\mu_{max}(\vartheta_{out}\leq
180^{\circ})\simeq 2.25 $, $\mu_{max}(\vartheta_{out}\leq 24^{\circ})\simeq
2.15 $, $\mu_{max}(\vartheta_{out}\leq 12^{\circ})\simeq 2.08 $, and
$\mu_{max}(\vartheta_{out}\leq 1^{\circ})\simeq 2.06 $. The enhancement
monotonically decreases with growing positron energy and approximately halves
at $\varepsilon \approx $~250~MeV. Thus, positron spectra from a crystal target
are softer at $L=9.0$~mm as well, and this property is much more pronounced in
comparison with $ L=2.2$~mm.

\begin{figure}[h]
  \centering
  \includegraphics[width=0.48\textwidth
  ]{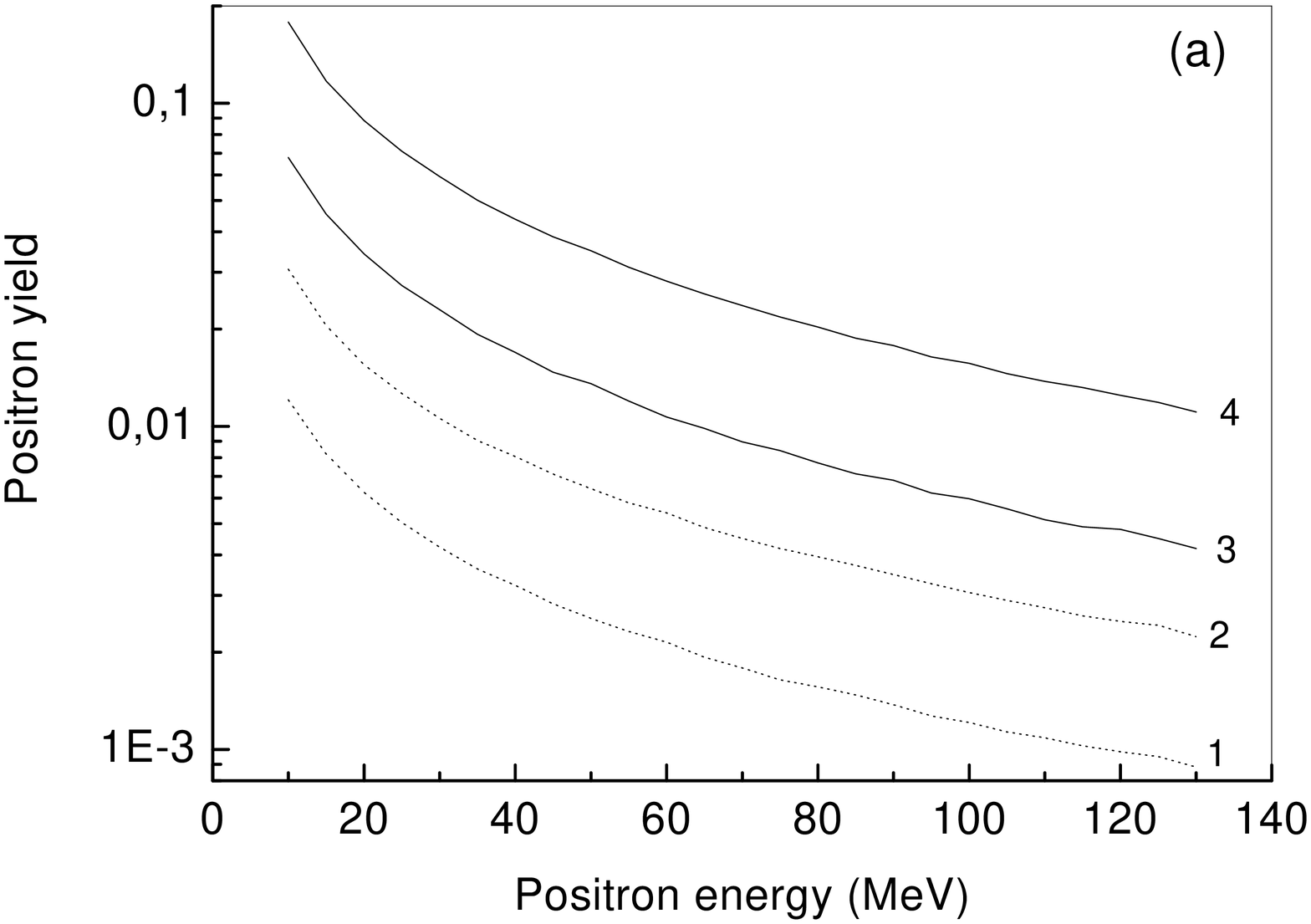}\hspace{0.025\textwidth}
 \includegraphics[width=0.48\textwidth
 ]{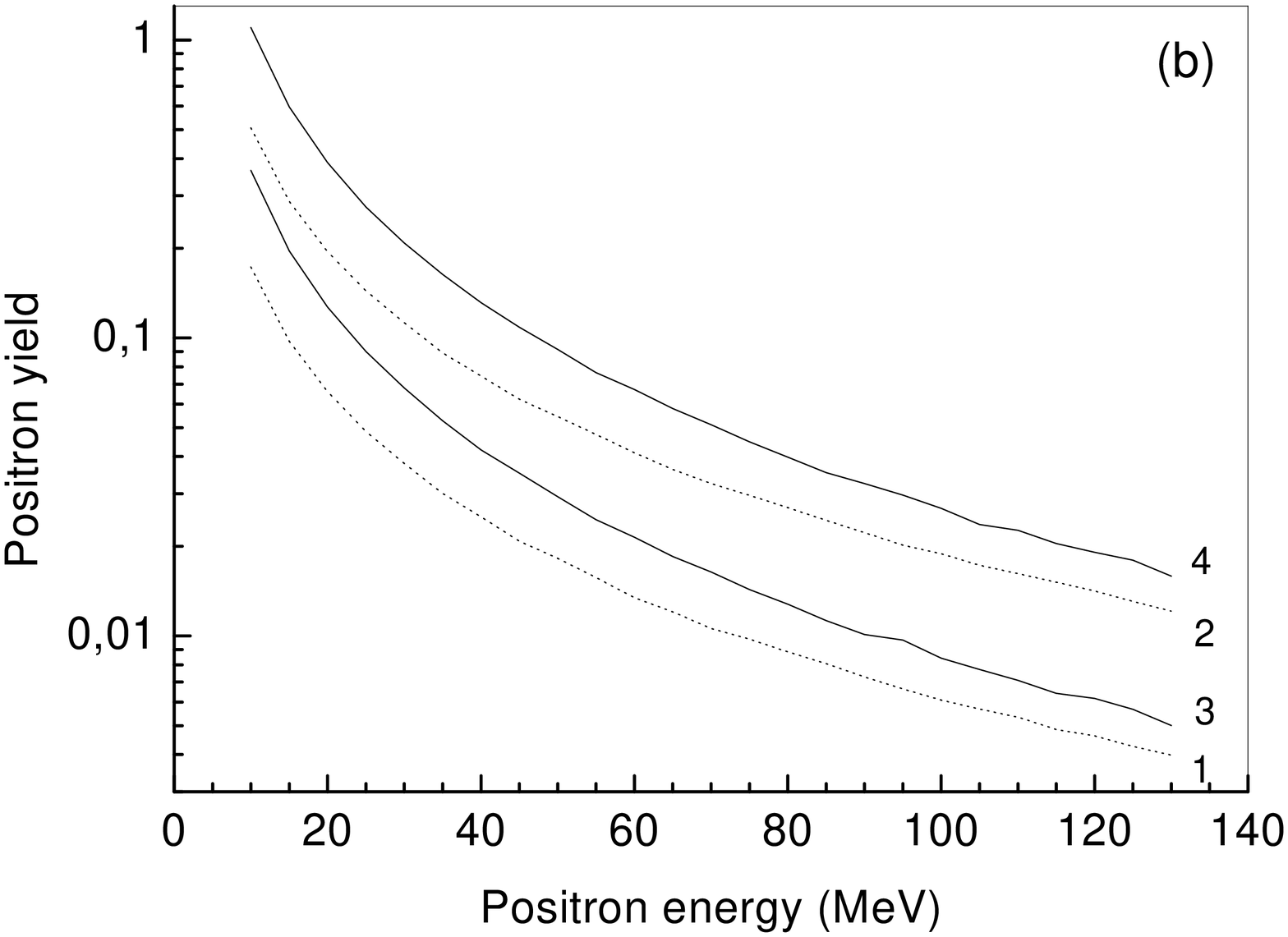}
 \caption{Positron yield  depending on energy at $L=2.2$~mm
 (a) and $L=9.0$~mm (b) for $p_{\perp}^{max}= $ 2.5~MeV/c ( curves 1 and
 3 ) and for $p_{\perp}^{max}= $ 5~MeV/c ( curves 2 and 4 ). Solid curves
 represent the yield from crystal and dotted from amorphous targets.}
 \label{Fig:Tran29}
 \end{figure}

\begin{figure}[h]
  \centering
  \includegraphics[width=0.48\textwidth
  ]{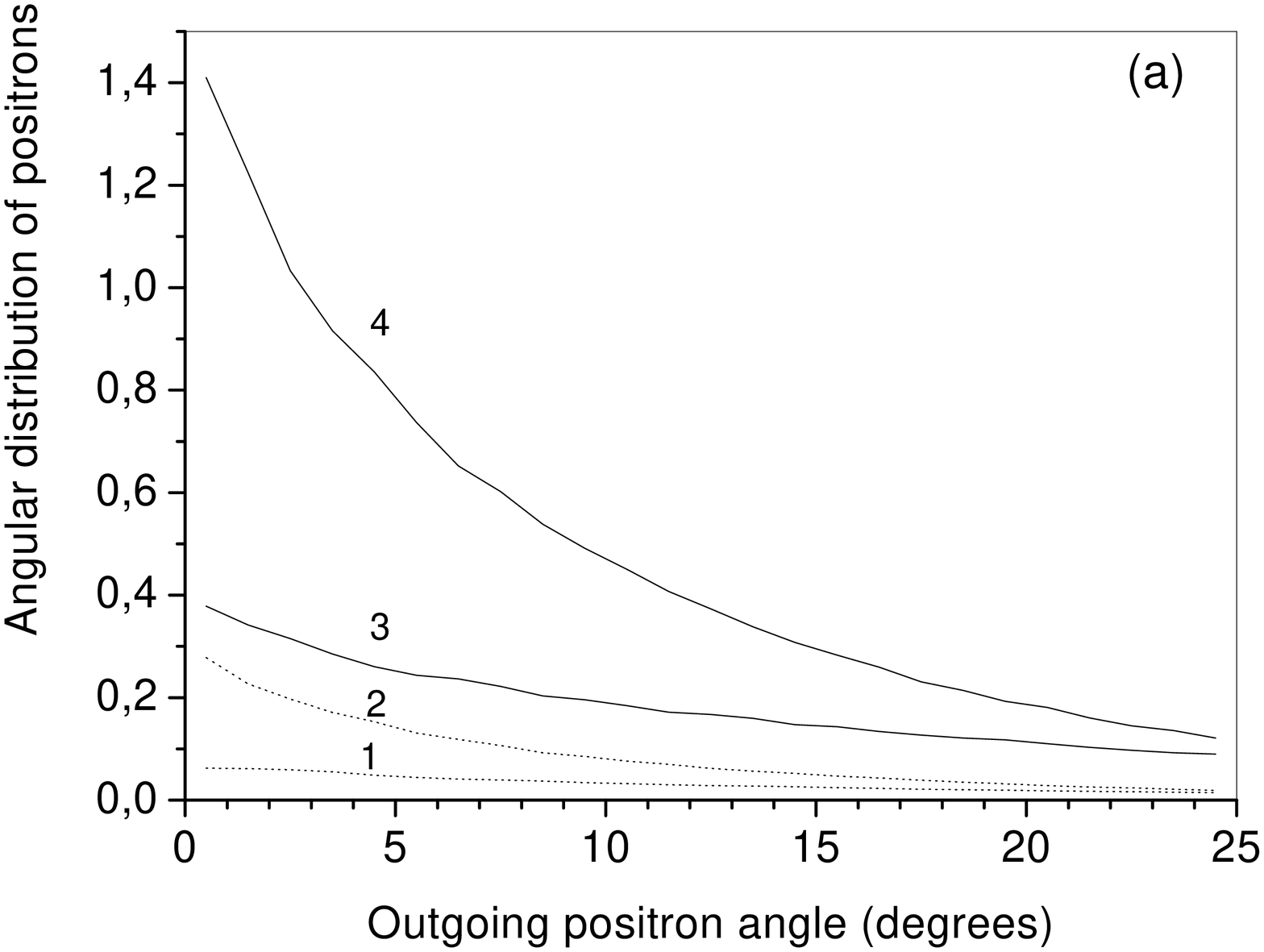}\hspace{0.025\textwidth}
 \includegraphics[width=0.48\textwidth
 ]{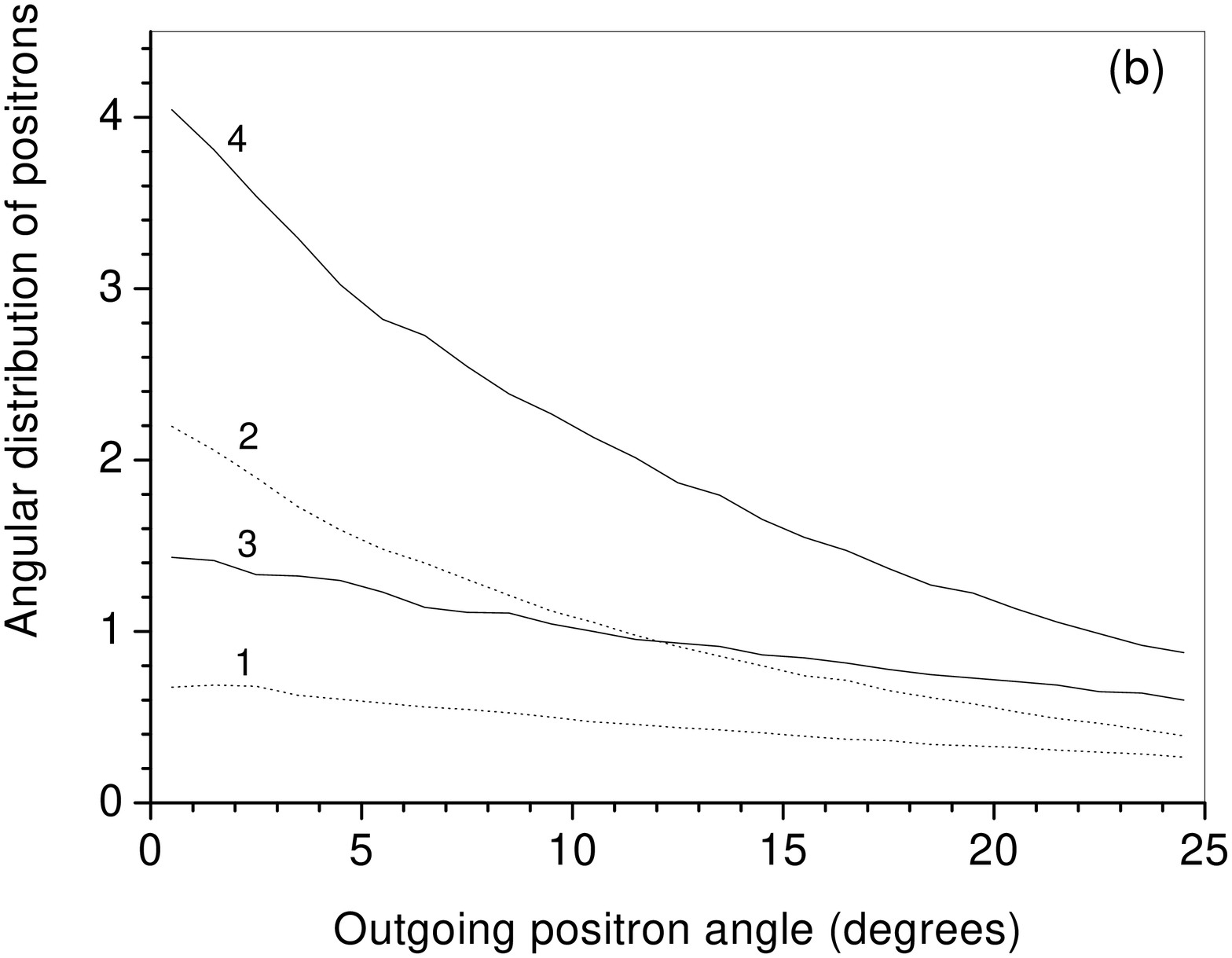}
 \caption{ Angular distribution $dN^{(+)}/d\Omega$ depending on outgoing
 positron angle at $L=2.2$~mm
 (a) and at $L=9.0$~mm (b) for $p \in (8.5 \div 11.5)$~Mev/c ( curves 1 and
3 ) and for $p \in (17 \div 23)$~Mev/c ( curves 2 and 4 ). Solid curves
 represent the yield from crystal and dotted from amorphous targets.}
 \label{Fig:Angu29}
 \end{figure}

Matching systems can be characterized also by a maximum transverse momentum
$p_{\perp}^{max} $ of accepted positrons. In this connection, spectra of
positrons having $p_{\perp} < p_{\perp}^{max} $ are of undoubted interest. Such
spectra at $L=2.2$~mm (a) and at  $L=9.0$~mm (b) from crystal and amorphous targets are
shown in Fig.\ref{Fig:Tran29}
In contrast to the case of the pure angular selection ( cf.
Figs.\ref{Fig:Spec2},\ref{Fig:Spec9}), the position of spectral maxima at
limited $p_{\perp}$ values is always in the first bin ( $\varepsilon \in (7.5
\div 12.5)$~MeV). Corresponding maximum values are $\mu_{max}(5~MeV/c)\simeq
5.82 $,  $\mu_{max}(2.5~MeV/c)\simeq 5.62 $ at $L=2.2$~mm and
$\mu_{max}(5~MeV/c)\simeq 2.17 $ , $\mu_{max}(2.5~MeV/c)\simeq 2.11 $ at
$L=9.0$~mm. The enhancement monotonically decreases with growing positron
energy. Its variation over the whole energy interval presented in
Fig.\ref{Fig:Tran29} is about 15~$\%$ at $L=2.2$~mm and 40~$\%$ at $L=9.0$~mm.
So, for this selection too, positron spectra from crystal targets are softer
than those from amorphous targets of the same thickness. The interesting
feature of spectral curves in Fig.\ref{Fig:Tran29} is the similarity of those
obtained for two different values of $p_{\perp}^{max}$ from the same target.
The scaling factors $\eta$ are $\eta_{cr} \simeq 2.6$, $\eta_{am} \simeq 2.5$
at $L=2.2$~mm and  $\eta_{cr} \simeq 3.1$, $\eta_{am} \simeq 3.0 $ at
$L=9.0$~mm. These factors turn out to be practically ( within an accuracy of a
few percent ) independent of the total positron momentum $p$. This fact can be
easily understood if we assume that a width of the angular distribution of
positrons is completely due to multiple scattering being, thereby, proportional
to $p^{-1}$. Such assumption is confirmed by results of the calculation shown
in Fig.\ref{Fig:Angu29} for two groups of positrons. One of them contains
positrons having momentum in the interval $ p \in (8.5 \div 11.5)$~Mev/c , for
another group $ p \in (17 \div 23)$~Mev/c.
For a given target, a width of the angular distribution of positrons with $p
\approx  10$~Mev/c is approximately twice as much that for $p \approx 20$~Mev/c
as expected. The width of every distribution evidently increases when we go on
to the thicker target of the same kind. Comparing angular distributions from
crystal and amorphous targets of the same thickness, we find that at $L=9.0$~mm
the distributions are somewhat ( about $ 1.5^{\circ}$ ) wider in the crystal
case for both groups. In units of FWHM of the distribution from the crystal
target these differences are about 6.5~$\%$ at $p \approx 10$~Mev/c and 14~$\%$
at $p \approx 20$~Mev/c. At $L=2.2$~mm the distribution from the crystal target
is wider by 15.5~$\%$ at $p \approx 20$~Mev/c whereas this is narrower by
10~$\%$ at $p \approx 10$~Mev/c.

Going on to a comparison of our results with those obtained in  \cite{O01}, let
us remind that to perform an accurate comparison of such kind, exact
information is needed concerning the acceptance conditions and registration
efficiency of detectors in the experiment. As noted in \cite{O01}, at $p =
20$~Mev/c, the momentum acceptance ($\Delta p/p$) was 3~$\%$ (FWHM) and the
polar angle acceptance was less than 20~mrad (FWHM). Since the shape of the
acceptance curves was unavailable to us, we have tried to simulate experimental
conditions using the same angular collimation $\vartheta_{out}\leq
\vartheta_{out}^{max}$ and the same value of $\Delta p/p$ for all momenta and
targets. So, at the calculation of the magnitudes of positron production
efficiency (PPE), we simply put $\vartheta_{out}^{max}$ to 20~mrad. The value
of $\Delta p/p$ was chosen to reproduce at applied collimation the experimental
magnitude of PPE for the 9.0~-~mm~-~thick amorphous target. Acting in this way,
we have got $\Delta p/p = 3.2$~$\%$.  We realize that our regard for the
acceptance conditions is rather rough. An additional inaccuracy was introduced
when we determined the PPE numbers from Fig.5 of \cite{O01}. Note that the
  \begin{figure}[h]
  \centering
  \includegraphics[width=0.48\textwidth
  ]{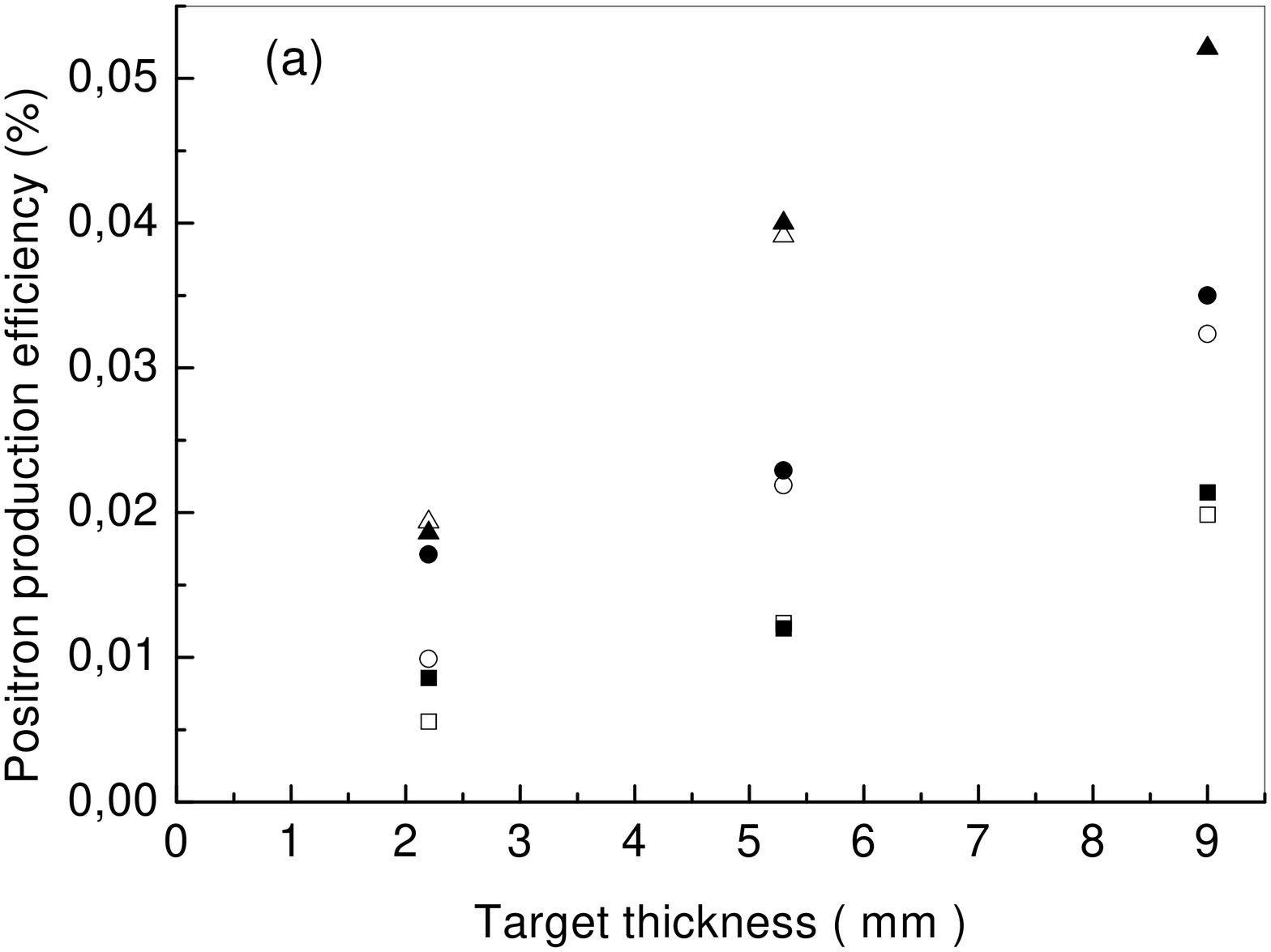}\hspace{0.025\textwidth}
 \includegraphics[width=0.48\textwidth
 ]{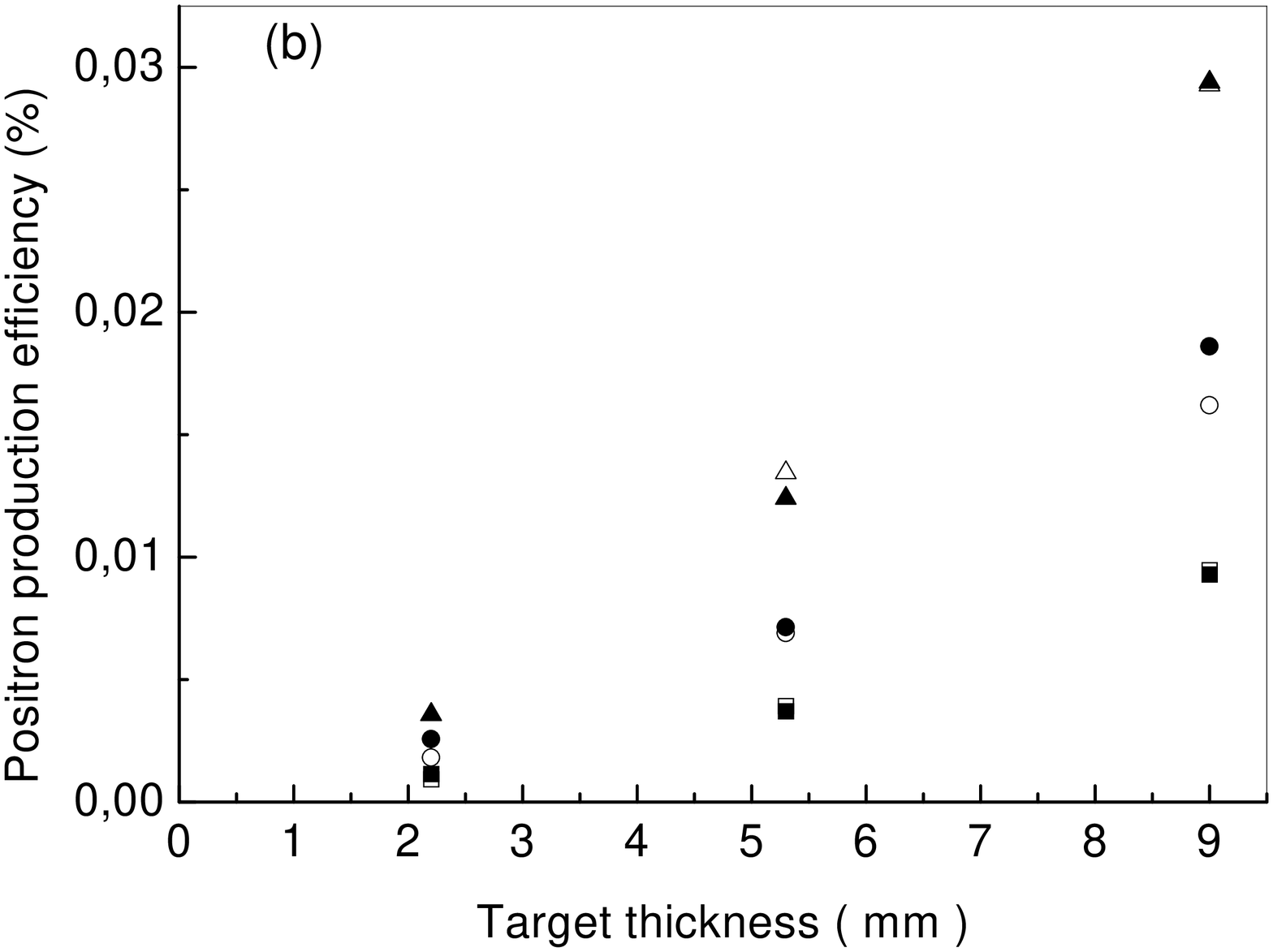}
 \caption{  Positron production efficiency from
 crystal (a) and  amorphous (b) targets depending on thickness.
 Open symbols - our calculation, filled symbols - results from Fig.5 of
 \cite{O01}; $\triangle$ are for $p = 20 $~Mev/c, $\bigcirc$ are for $p = 15 $~Mev/c,
  and $\square$ are for $p = 10 $~Mev/c.}
 \label{Fig:Abscram}
 \end{figure} 
experimental numbers obtained in such a way, which are presented by filled
symbols in Fig.\ref{Fig:Abscram}, do not reproduce exactly the whole set of
mean experimental values for the enhancement given in  Table 1 of
\cite{O01}. Moreover, in Fig.5 of \cite{O01} there are no experimental points
for 2.2 and 5.3~-~mm~-~thick amorphous targets. In these cases the values of
PPE given by smooth~-~curve fits presented  in Fig.5 of \cite{O01}  were used
by us as experimental results. Bearing all this in mind, we, nevertheless, can
assert that a rather good agreement is seen in Fig.\ref{Fig:Abscram} of the
experimental results and our estimations. Relative difference of them is better
than 13~$\%$ everywhere except the values of PPE at $p =$ 10 and 15~Mev/c from
both thinnest ( $L=2.2$~mm ) targets, where the experimental yield is
underestimated by 19~$\%$ to 42~$\%$. Note that just for this thickness the
largest inaccuracy was introduced while determining the PPE numbers from Fig.5
of \cite{O01} at $p =$ 10 and 15~Mev/c, as the magnitude of the yield is
especially small in this case.

\begin{table}[h]
\begin{center}
\caption{ \label{Tab:Enhan} Enhancement of the positron yield from crystal targets}
\vskip .5cm
\begin{tabular}{|c|c|c|c|c|c|c|}   \hline
Momentum & \multicolumn{2}{|c|}{Enhancement}& \multicolumn{2}{|c|}{Enhancement}
& \multicolumn{2}{|c|}{Enhancement} \\ ( MeV/c )& \multicolumn{2}{|c|}{(
2.2-mm-thick)}& \multicolumn{2}{|c|}{( 5.3-mm-thick)} & \multicolumn{2}{|c|}{(
9.0-mm-thick)} \\ \cline {2-7}& theory & experiment & theory & experiment&
theory & experiment \\ \hline 10 & $6.0 \pm 0.5$ & $6.5 \pm 0.6$ & $3.2 \pm
0.3$ & $3.4 \pm 0.7$ & $2.1 \pm 0.2$ & $2.3 \pm 0.4$
\\ \hline 15 & $5.5 \pm 0.3$ & $6.2 \pm 0.8$ & $3.2 \pm
0.2$ & $3.2 \pm 0.5$ & $2.0 \pm 0.1$ & $2.0 \pm 0.2$
\\ \hline 20 & $5.4 \pm 0.2$ & $5.1 \pm 0.5$ & $2.9 \pm
0.1$ & $3.0\pm 0.5$ & $1.8 \pm 0.1$ & $1.8 \pm 0.2$ \\ \hline
\end{tabular}
\end{center}
\end{table}

In contrast to the magnitude of the positron yield, the enhancement is not very
sensitive to the acceptance conditions.  The
calculated  values of the enhancement( theory ) are presented in Table \ref{Tab:Enhan} along with those taken from  Table 1 of \cite{O01} ( experiment ). Purely statistical errors
are figured in  Table \ref{Tab:Enhan} as theoretical ones. The relative
error in PPE was estimated as $N_{ef}^{-1/2}$, where $N_{ef}$  is the mean number
of events in the phase space corresponding to the acceptance conditions used in
calculations. The total statistics was chosen so that approximately to equalize
values of $N_{ef}$ for amorphous and crystal targets of the same thickness. At
given  total statistics, the quantity  $N_{ef}$ increases with growing positron
momentum in accord with a shape of the positron spectra at hard collimation
shown in Figs. \ref{Fig:Spec2},\ref{Fig:Spec9}. This fact leads to a better
statistical accuracy for larger momentum. We emphasize that the differences of
the estimated and experimental enhancement values are smaller than
corresponding experimental errors for all momenta and samples figured in
Table \ref{Tab:Enhan}.
\section{Conclusion}

Using a simple computer code suggested in \cite{BKS4} and  \cite{BS1}, we have
compared the theoretical predictions for some characteristics of the
electromagnetic shower developing in axially aligned crystals with experimental
results reported in \cite{CBA00},\cite{C01} and \cite{ITY00},\cite{O01}. On the
whole, theory and experiment are consistent within an experimental accuracy.
From this comparison we also conclude that the accuracy provided by the
existing simplified code is at least better than 20$\%$. This accuracy may be
slightly improved if we include into consideration  some processes like
annihilation of positrons or Compton scattering of photons which were ignored
as corresponding cross sections are small in the energy region of interest.
However, the approximate character of the radiation spectra at axial alignment
used in our calculations still provides the main  theoretical uncertainty.
Nevertheless, we believe that a level of the accuracy already achieved in the
theoretical description is quite sufficient to make a reliable choice for optimal
parameters of the positron source using axially aligned single crystals.

\vspace{0.25 cm}
{\bf Acknowledgements}
\vspace{0.25 cm}

We are grateful to Prof. H.Okuno for providing us with details of the
experiment \cite{O01} and to the authors of \cite{C01} for numerous fruitful
discussions.  Support of this work by the Russian Fund of Basic Research under
Grants 00-02-18007,  01-02-16926, and 01-02-22003 is also gratefully
acknowledged.


\end{document}